\documentclass[aps,prl,twocolumn,amsmath,amssymb,superscriptaddress]{revtex4}
\usepackage{graphicx}
\usepackage{subfigure}
\usepackage{epsfig}
\usepackage{bm}
\usepackage{ulem}
\usepackage{color}
\usepackage{multirow}
\usepackage{slashed}
\usepackage{hyperref}
\usepackage{verbatim}
\usepackage{tikz}
\usepackage{pgffor}						
\usepackage{hyperref}

\begin{document}

\title{Non-Hermitian quantum system generated from two coupled Sachdev-Ye-Kitaev models}
\author{Wenhe Cai}
\affiliation{Department of Physics, Shanghai University, Shanghai, 200444, China}
\author{Sizheng Cao}
\affiliation{Department of Physics, Shanghai University, Shanghai, 200444, China}
\author{Xian-Hui Ge}
\email{gexh@shu.edu.cn}
\affiliation{Department of Physics, Shanghai University, Shanghai, 200444, China}
\author{Masataka Matsumoto}
\affiliation{Department of Physics, Shanghai University, Shanghai, 200444, China}
\author{Sang-Jin Sin}
\email{sjsin@hanyang.ac.kr}
\affiliation{Department of Physics, Hanyang University, Seoul 04763, Korea}

\begin{abstract}
We show that a non-Hermitian two coupled Sachdev-Ye-Kitaev (SYK) model can provide thermodynamic structure equivalent to Hermitian two coupled SYK model. The energy spectrum, the entanglement degree of the ground states and the low energy effective action of this model are not influenced by the non-Hermiticity. The novel biorthogonal ground states demonstrates that two SYK sites, one of which can be in the ground state and the other in the Schwarzian excited state by tuning the non-Hermiticity. We find evidence that the free energy is independent of the non-Hermiticity.
\end{abstract}
\date{\today}
\maketitle

\section{I.Introduction}
Recently, non-Hermitian physics has gained widespread attention, such as non-Hermitian linear response theory \cite{PCCZ}, non-Hermitian topological systems \cite{SZF,Ge:2019crj} and non-Hermitian holography \cite{Arean:2019pom,Liu:2020abb}. One of the primary reasons for those studies is due to the probability in nature effectively becomes non-conserving due to the presence of energy, particles, and information regarding the external degrees of freedom that are out of the Hilbert space.  In a non-Hermitian system experiencing an exceptional point in the wave momentum, the corresponding eigenfrequencies change from real to complex numbers \cite{Bender:2007nj,NHOPTICS1,NHOPTICS2012,NHOPTICS18,NHOPTICS17}.
However, seminal work by C. Bender and S. Boettcher demonstrates that in the physics of non-Hermitian systems, a huge class of non-conservative Hamiltonians can exhibit entirely real spectra as long as they commute with the parity-time (PT) operator \cite{Bender:1998ke}.
Moreover, all the PT symmetric Hamiltonians studied reported in the literature exhibited such property \cite{mostafa1,mostafa2,mostafa3}. The similarity transformations can also enable one to construct non-Hermitian Hamiltonian with real spectrum \cite{mostafa3,Zhao:2020xrt, Zhao:2020khc}.

In a quantum many-body system, the quantum two-level system can be simulated by coupling two copies of the Sachdev-Ye-Kitaev (SYK) model. The SYK mode, which is well-known as a disordered and strongly-coupled quantum system composed of Majorana fermions \cite{SY,K,Maldacena:2016upp}, has recently emerged as an exemplary model providing insight into the nature of non-Fermi liquids \cite{Phillips:2019qva}, quantum chaos \cite{Jensen:2016pah}, holography \cite{Maldacena:2016hyu,He:2021dhr}, strange metallic transport \cite{Sachdev:2015efa,Ge:2018lzo} and high temperature superconducting \cite{Cai:2018lsr,Salvati:2021eos}. The SYK model is closely related to two-dimensional dilaton gravity describing excitations above the near horizon external black hole \cite{almheiri14,mertens16}. Therefore, an eternal traversable wormhole can be constructed by considering two copies of SYK models coupled by a simple interaction. This Maldacena-Qi (MQ) model demonstrates that at low temperature, the coupling can drive phase transitions to a phase holographically dual to an eternal traversable wormhole with an $AdS_2$ throat \cite{Maldacena:2018lmt}.  Conversely, at a higher temperature, the system reduces to two gapless black hole phases \cite{Maldacena:2018lmt}. However, in non-Hermitian set up, one may expect this gapped-gapless physical picture drastically changed.

\begin{figure}[!t]
\label{figureone}
\centerline{\includegraphics[width=7.0cm]{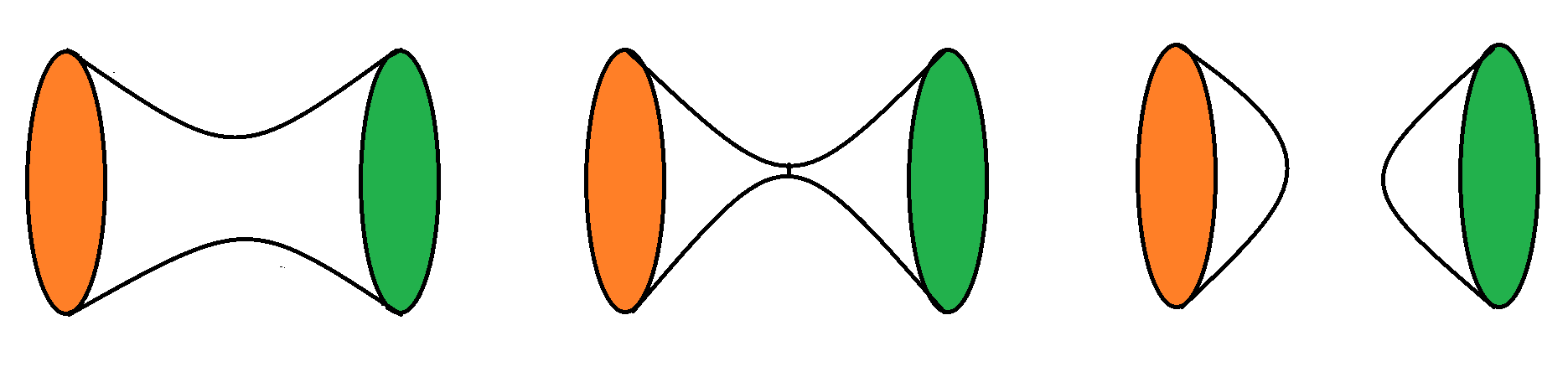}}
\caption{\label{fig:Figure phase} Sketched phase diagram for the physical picture. Left: A geometry connecting the two sides in the low temperature regime. The left Sachdev-Ye-Kitaev (SYK) and right SYK are in different states because of the non-Hermitian parameter.
Middle: An unstable geometry connecting two SYK sites. Right: Two separated SYK sites at high temperature represent the gapless two black hole phases. The non-Hermitian parameter can change the states of the left and right SYK sites, which are marked in orange and green colors. When $\alpha=0$, the phase diagram recovers that of \cite{Maldacena:2018lmt}.}
\end{figure}

The main goal of this paper is to prove that the wormhole-black hole picture is robust not only in a Hermitian two coupled SYK model, but also in a non-Hermitian two coupled SYK model. For this purpose, we propose a novel non-Hermitian two-site SYK model. We first prove that the system yields real energy spectrum. Furthermore, we show that the degree of entanglement, the low energy effective action and the phase structure are non-Hermiticity independent.
As illustrated in Fig.\ref{fig:Figure phase}, even though these two SYK sites are approaching a ``ground state/excited state" picture in the regime of strong non-Hermitian limit, the thermodynamic phase structure indicates three distinct properties at different temperatures.

\section{II.Non-Hermitian two coupled SYK model}
We consider non-Hermitian two coupled SYK model with the Hamiltonians
\begin{align}
H&=-\sum^N_{ijkl}J_{ijkl}\sum_{A=R,L}(c_1 C^{A\dag}_iC^{A\dag}_jC^{A}_kC^{A}_l\nonumber\\
 &+ c_2 C^{A\dag}_iC^{A}_jC^{A\dag}_kC^{A}_l)+H_{int},\nonumber\\
H_{\rm int}&=i\mu\sum^N_i(e^{-2\alpha}C^{L\dag}_iC^{R}_i-e^{2\alpha}C^{R\dag}_iC^{L}_i),
\end{align}
where $A=L,R$ refers to the ``left" and ``right" side of the two identical copies, and $c_1$ and $c_2$ are two real valued constants. We choose $c_1=2$ and $c_2=4$ in what follows to match to the MQ model \cite{Maldacena:2018lmt}. Note that other choices of positive $c_1$ and $c_2$ do not change the total physical picture. The parameter $\alpha$ is a real number controlling the strength of non-Hermiticity, which is introduced by a non-Hermitican particle-hole similarity transformation (see appendix A for details). Under the self-similarity transformation, Dirac fermions $C$ and $C^\dagger$ are introduced to replace the Majorana fermions $\psi$ in the MQ model. The coupling $J_{ijkl}$ are random real numbers, which obey the Gaussian distribution and satisfy $J_{ijkl}=-J_{jikl}=-J_{ijlk}=J_{klij}$ with
$<J_{ijkl}>=0\,,~\,<J^2_{ijkl}>=\frac{J^2}{8N^3}\,.$ For this reason, we call this model a pseudo-complex SYK to distinguish from the usual complex SYK. By analytically continued to an imaginary value $\alpha\rightarrow i\alpha$, one can recover the Hermitian Hamiltonian.

\section{III.Energy spectrum and degree of entanglement}
\begin{figure}[!t]
\begin{minipage}{0.48\linewidth}
\centerline{\includegraphics[width=4.5cm]{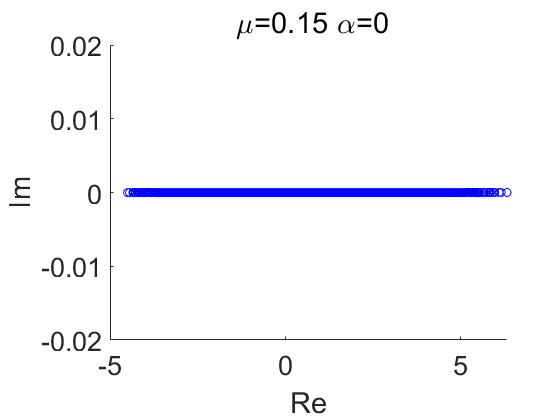}}
\centerline{(a)}
\end{minipage}
\hfill
\begin{minipage}{0.48\linewidth}
\centerline{\includegraphics[width=4.5cm]{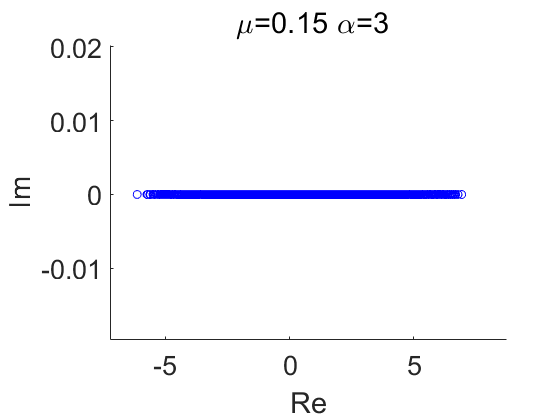}}
\centerline{(b)}
\end{minipage}
\caption{\label{fig:Figure 2} Plots of the spectrum with $\mu=0.15,N=8$ for the values of $\alpha=0$ (a) and for the values of $\alpha=3$ (b).}
\end{figure}

Energy spectrum is an important feature of the non-Hermitian quantum system. We compute the energy spectrum by using exact diagonalization techniques. The energy spectrum is real and independent of the non-Hermitian parameter $\alpha$, as demonstrated in  Fig. \ref{fig:Figure 2}.

  For non-Hermitian systems, we need  construct the ground state by introducing a biorthogonal set $\{|\psi^l_n\rangle,|\psi^r_n\rangle\}$ \cite{mostafa3,Zhao:2020xrt,Zhao:2020khc,Matzkin:2006}.
  The right/left eigenstates are defined as
\begin{equation}
  H|\psi^r_n\rangle=E_n|\psi^r_n\rangle\,,\,H^\dag|\psi^l_m\rangle=E^*_m|\psi^l_m\rangle\,.
\end{equation}
The eigenstates satisfy the properties as follow
\begin{equation}
  \sum_n|\psi^l_n\rangle\langle\psi^r_n|=I\,,\,\langle\psi^l_n|\psi^r_m\rangle=\delta_{nm}\,.
\end{equation}
We impose the constraints that system  yields the ground state energy as those of \cite{Sahoo:2020unu} by taking  $H^\dag_{int}|\psi_0^l\rangle=-\mu N|\psi_0^l\rangle$, $H_{int}|\psi_0^r\rangle=-\mu N|\psi_0^r\rangle$ and $\langle \psi_0^l|\psi_0^r\rangle=1$.
Without loss of generality, the generated ground states are proposed as
\begin{eqnarray}
  |\psi_0^r\rangle&=&\prod^N_{j}(\tilde{A}|1\rangle_{L,j}|0\rangle_{R,j}+i \tilde{B}|0\rangle_{L,j}|1\rangle_{R,j})\,,\label{eq:gs1}\\
  |\psi_0^l\rangle&=&\prod^N_{j}(\tilde{C}|1\rangle_{L,j}|0\rangle_{R,j}+i \tilde{D}|0\rangle_{L,j}|1\rangle_{R,j})\,,\label{eq:gs2}
\end{eqnarray}
where the coefficients $\tilde{A}, \tilde{B}, \tilde{C}$ and $\tilde{D}$ satisfy the relation $\tilde{B}=\tilde{A} e^{2\alpha}$ and $\tilde{D}=\tilde{C} e^{-2\alpha}$.
The constraint $\langle \psi_0^l|\psi_0^r\rangle=1$ further leads to $\tilde{A}  \tilde{C}=\tilde{B}  \tilde{D}=\frac{1}{2}$. The ground states can return to the ground state of Hermitian case ($\alpha=0$) consistently if setting
\begin{equation}
\tilde{A}=\frac{1}{\sqrt{2}}\,,\,\tilde{B}=\frac{e^{2\alpha}}{\sqrt{2}}\,,\,\tilde{C}=\frac{1}{\sqrt{2}}\,,\,\tilde{D}=\frac{e^{-2\alpha}}{\sqrt{2}}\,.
\end{equation}
In the biorthogonal set, the usual orthogonal normalization is not applicable.

In our non-Hermitian model, the degree of entanglement turns out to be
\begin{align}
P_E&=-tr(\rho_A\log \rho_A)=1\,,
\end{align}
which can be derived from the von-Neumann entropy of the reduced density matrix
\begin{align}
\rho_{LR}&=|\psi_0^r\rangle\langle \psi_0^l|=\tilde{A}  \tilde{C}|10\rangle\langle 01|+i \tilde{B}  \tilde{C}|01\rangle\langle 01|\nonumber\\
&-i\tilde{A}  \tilde{D}|10\rangle\langle 10|+\tilde{B}  \tilde{D}|01\rangle\langle 10|)\,.\\
\rho_{L}&=tr_R(\rho_{LR})=\left(
                                                   \begin{array}{cc}
                                                     \tilde{A}  \tilde{C} & 0 \\
                                                     0 & \tilde{B}  \tilde{D} \\
                                                   \end{array}
                                                 \right)
\,.
\end{align}
Because of $P_E=1$, which is independent of $\alpha$, the ground states are maximally entangled between two systems. The ground states (\ref{eq:gs1}) and (\ref{eq:gs2}) will approach the pure state for large $\alpha$, but the degree of entanglement remains unchanged. Taking the limit $\alpha\rightarrow -\infty$, we have
\begin{equation}
 |\psi_0^r\rangle\rightarrow\prod^N_{j}\frac{1}{\sqrt{2}}|1\rangle_{L,j}|0\rangle_{R,j}\,,\,|\psi_0^l\rangle\rightarrow\prod^N_{j}\frac{i e^{-2\alpha}}{\sqrt{2}}|0\rangle_{L,j}|1\rangle_{R,j}\,,
\end{equation}
while in the limit $\alpha\rightarrow +\infty$, we obtain
\begin{equation}
 |\psi_0^l\rangle\rightarrow\prod^N_{j}\frac{i}{\sqrt{2}}|0\rangle_{L,j}|1\rangle_{R,j}\,,\,|\psi_0^r\rangle\rightarrow\prod^N_{j}\frac{i e^{2\alpha}}{\sqrt{2}}|0\rangle_{L,j}|1\rangle_{R,j}\,.
\end{equation}
Therefore, the left and right SYK sites are no longer symmetric as illustrated in Fig.\ref{fig:Figure phase}.

\section{IV.Low energy effective action}
In order to see the effects of the non-Hermitian parameter, we study low energy properties of both in the extended SYK model and the gravity description. In the low energy limit, the model simplifies due to the emergence of a conformal symmetry.
The retarded Green's function of non-Hermitian system is defined as
\begin{equation}
  G_{AB}(\tau_1,\tau_2)=\frac{1}{N}\sum_n\langle\psi^l_n|TC^{A\dag}_i(\tau_1)C^{B}_i(\tau_2)|\psi^r_n\rangle\,,
\end{equation}
where $A,B=L,R$.
The saddle-point equations are invariant under the time reparametrization $\tau\rightarrow h(\tau)$ and the $U(1)$ symmetry, which is the same as the complex SYK model in \cite{Cai:2017vyk,Davison:2016ngz,Jian:2017unn,Zhang:2020szi}.
\begin{align}
\tilde G_{AB}(\tau_1,\tau_2)&=[h_A^\prime(\tau_1)h_B^\prime(\tau_2)]^{\Delta}G_{AB}\big(h(\tau_1),h(\tau_2)\big)\nonumber\\
                             &e^{i\phi_A(\tau_1)-i\phi_B(\tau_2)}\,,\nonumber\\
\tilde \Sigma_{AB}(\tau_1,\tau_2)&=[h_A^\prime(\tau_1)h_B^\prime(\tau_2)]^{1-\Delta}\Sigma_{AB}\big(h(\tau_1),h(\tau_2)\big)\nonumber\\
                                 &e^{i\phi_B(\tau_2)-i\phi_A(\tau_1)}\,,\nonumber
\end{align}
 In the absence of the interacting term, the Schwarzian effective action of the left or right copy turns out to be
\begin{align}
 S_{A}&=-N\alpha_S\int d\tau\{\tanh\frac{h_A(\tau)}{2},\tau\}\nonumber\\
      &+\frac{NK}{2}\int d\tau\bigg(\phi^\prime_A(\tau)+i\varepsilon_A h^\prime_A(\tau)\bigg)^2\,.
      \label{eq:SA}
\end{align}
where $\varepsilon_A$ is related to the the $U(1)$ charge $Q_A$ and $\alpha_S$ is determined by four-point calculation of the SYK model. The above effective action is given by the Schwarzian derivative,
\begin{equation}
 \{h,\tau\}=\frac{h^{\prime\prime\prime}(\tau)}{h^\prime(\tau)}-\frac{3}{2}\bigg(\frac{h^{\prime\prime}(\tau)}{h^\prime(\tau)}\bigg)^2\,.\nonumber
\end{equation}
The effective action of the coupled part is written as
\begin{align}
 S_{int}&=\frac{\mu}{2}\int d\tau\bigg[\frac{bh^\prime_L(\tau)h^\prime_R(\tau)}{\cosh^2\frac{h_L(\tau)-h_R(\tau)}{2}}\bigg]^\Delta\nonumber\\
 &\cosh(\varepsilon h_L(\tau)-\varepsilon h_R(\tau))\nonumber\\
 &[e^{i(\phi_L-\phi_R)-2\alpha}+e^{-i(\phi_L-\phi_R)+2\alpha}]\,.\label{eq:Sint1}
\end{align}
This action has the global $SL(2)\times U(1)$ symmetry generated by
\begin{align}
&\delta h_L=\epsilon^0+\epsilon^+e^{ih_L}+\epsilon^-e^{-ih_L}\,,\nonumber\\
&\delta h_R=\epsilon^0-\epsilon^+e^{ih_R}-\epsilon^-e^{-ih_R}\,,\nonumber\\
&\delta\phi_A=-i\varepsilon\delta h_A+\epsilon\,,
\label{eq:generator}
\end{align}
where $\varepsilon_L= \varepsilon_R = \varepsilon$.

 The total action could be simplified to
\begin{align}
 \frac{S}{N}&=-2\alpha_S\int d\tau\{\tanh\frac{h(\tau)}{2},\tau\}+K\int d\tau\bigg(\phi^\prime(\tau)+i\varepsilon h^\prime(\tau)\bigg)^2\nonumber\\
 &+\frac{\mu}{2^{2(\Delta-1)}}\int d\tau\big(h^\prime(\tau)\big)^{2\Delta}\,,
\end{align}
with the solution
\begin{equation}
 h_L=h_R=h(\tau)\,,\,\phi_L=\phi_R-2i\alpha=\phi(\tau)\,,\label{sol}
\end{equation}
where we have chosen $b^\Delta=N$. In the appendix B, we derive the $SL(2)$ Noether charges. The $SL(2)$ Noether charge vanishes with the simple solution $h(\tau)=t^\prime\tau$, so it can be treated as gauge symmetry. The solutions of equation (\ref{sol}) lead  to $Q_{\pm}=0$, and
\begin{equation}
 Q_0/N=2e^{-\phi}[-\phi^{\prime\prime}-e^{2\phi}+\Delta\mu e^{2\Delta\phi}]=0\,,
\end{equation}
by introducing $\phi=\log h^\prime$. We can derive the equations of motion from the action of a non-relativistic particle in a potential
\begin{equation}
 S=N\int du\bigg[(\phi^\prime)^2-\big(e^{2\phi}-\mu e^{\phi/2}\big)\bigg]\,.
\end{equation}
The effective potential is independent of the non-Hermitian parameter $\alpha$, but same as that of MQ model \cite{Maldacena:2018lmt}. One can therefore conclude that there is an $\alpha$-independent energy gap at low energy.

 We can thus add a boundary interaction to the bulk action
\begin{equation}
 S_{int}=g\sum^{N}_{i=1}\int du\bigg(e^{-2\alpha}O^{i\dag}_L(u)O^i_R(u)-e^{2\alpha}O^{i\dag}_R(u)O^i_L(u)\bigg)\,,\label{eq:Sint2}
\end{equation}
where $O$ is a set of $N$ operators with dimension $\Delta$ and $g$ is proportional to the coupling $\mu$. When $\alpha$ and $g$ is small, the interacting term in Eq.(\ref{eq:Sint2}) corresponds to the interaction term of the low energy effective action in Eq.(\ref{eq:Sint1}). The coupling of left and right black holes $(1-2\alpha)O^{i\dag}_LO^{i}_R\,,\,(1+2\alpha)O^{i\dag}_RO^{i}_L$ are not symmetric. The two sides of $AdS_2$ is directly coupled by the double trace deformation. Since $e^{2\alpha}$ or $e^{-2\alpha}$ is always positive, the double trace interaction generates negative null energy in the bulk without violating causality same as that of \cite{Gao:2016bin}. Therefore quantum entangled states at left and right boundaries are connected.

\section{V.Thermodynamic phase structure beyond the low energy limit}
At finite temperature, the retarded Green's function receives great contribution from the non-Hermitian parameter $\alpha$. But the whole thermodynamic phase structure remains unchanged as the non-Hermitian parameter $\alpha$ varies.

The effective action can be obtained as,
\begin{align}
\frac{S_{eff}}{N}&=-\log\det(\sigma_{AB}-\Sigma_{AB})-\int d\tau_1d\tau_2\bigg(\Sigma_{BA}(\tau_2,\tau_1)\nonumber\\
                 &G_{AB}(\tau_1,\tau_2)+\frac{36}{4}J^2G^2_{AB}(\tau_1,\tau_2)G^2_{BA}(\tau_2,\tau_1)\bigg), \label{eq:Seff}
\end{align}
where
\begin{equation}
  \sigma_{AB}=\begin{pmatrix}
                \partial_\tau & i\mu e^{-2\alpha} \\
                -i\mu e^{2\alpha} & \partial_\tau
               \end{pmatrix}.
\end{equation}
The corresponding equations can be written as,
\begin{align}
  &\Sigma_{AB}(\tau_1,\tau_2)=-36J^2G^2_{AB}(\tau_1,\tau_2)G_{BA}(\tau_2,\tau_1)\,,\nonumber\\
&G_{LL}(i\omega_n,\alpha)=\frac{-i\omega_n-\Sigma_{LL}(i\omega_n,\alpha)}{D(i\omega_n,\alpha)}\,,\nonumber\\
&G_{RR}(i\omega_n,\alpha)=\frac{-i\omega_n-\Sigma_{RR}(i\omega_n,\alpha)}{D(i\omega_n,\alpha)}\,,\nonumber\\
&G_{LR}(i\omega_n,\alpha)=\frac{-i\mu e^{-2\alpha}+\Sigma_{LR}(i\omega_n,\alpha)}{D(i\omega_n,\alpha)}\,,\nonumber\\
&G_{RL}(i\omega_n,\alpha)=\frac{i\mu e^{2\alpha}+\Sigma_{RL}(i\omega_n,\alpha)}{D(i\omega_n,\alpha)}\,,\nonumber\\
&D(i\omega_n,\alpha)=\bigg(-i\omega_n-\Sigma_{LL}\bigg)\bigg(-i\omega_n-\Sigma_{RR}\bigg)\nonumber\\
&+\bigg(i\mu e^{-2\alpha}-\Sigma_{LR}\bigg)\bigg(i\mu e^{2\alpha}+\Sigma_{RL}\bigg)\,,\label{eq:SDeq}
\end{align}
with the Matsubara frequency $\omega_n=2\pi(n+\frac{1}{2})/\beta$.
\begin{figure}[!t]
\begin{minipage}{0.48\linewidth}
\centerline{\includegraphics[width=4.5cm]{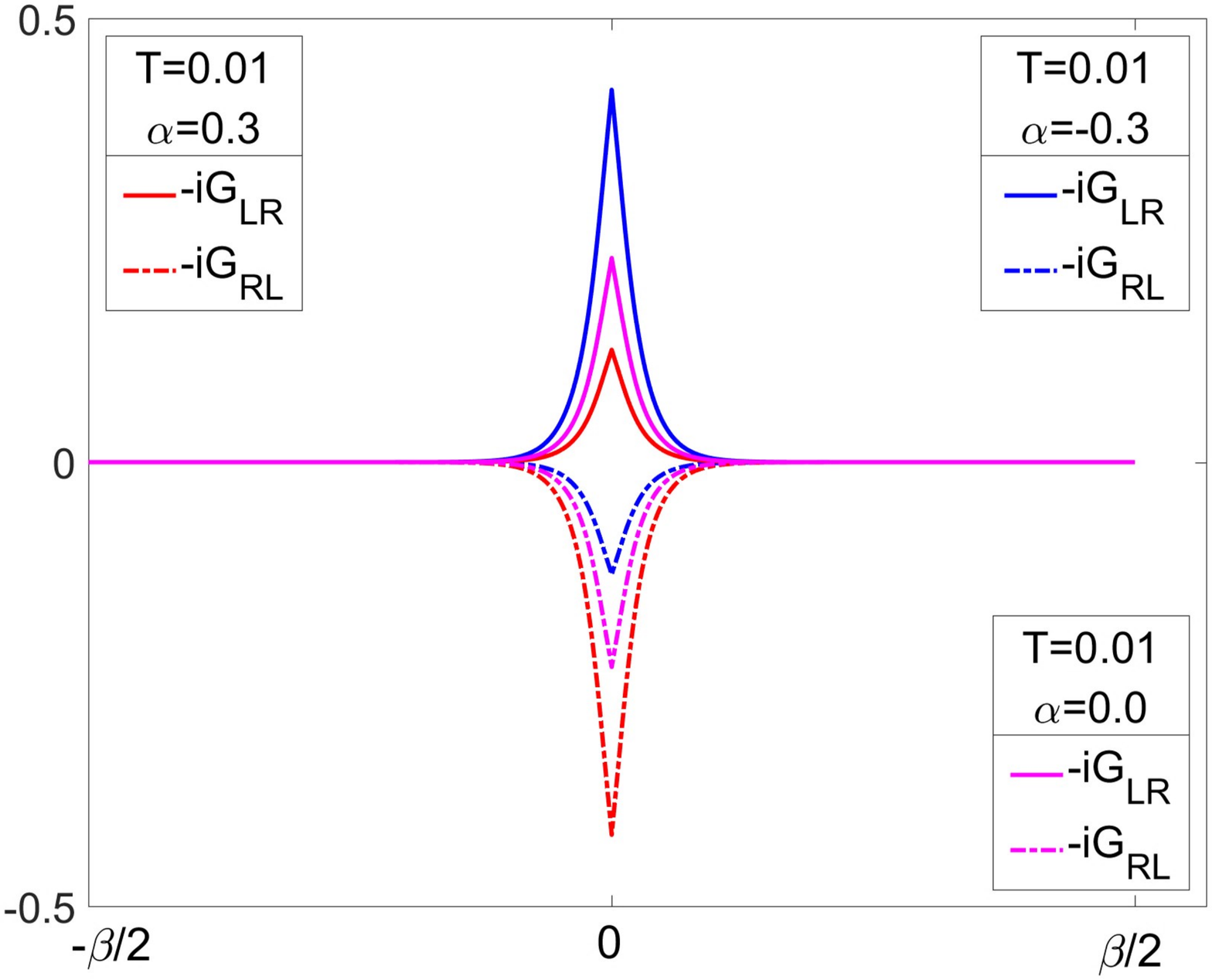}}
\centerline{(a)}
\end{minipage}
\hfill
\begin{minipage}{0.48\linewidth}
\centerline{\includegraphics[width=4.5cm]{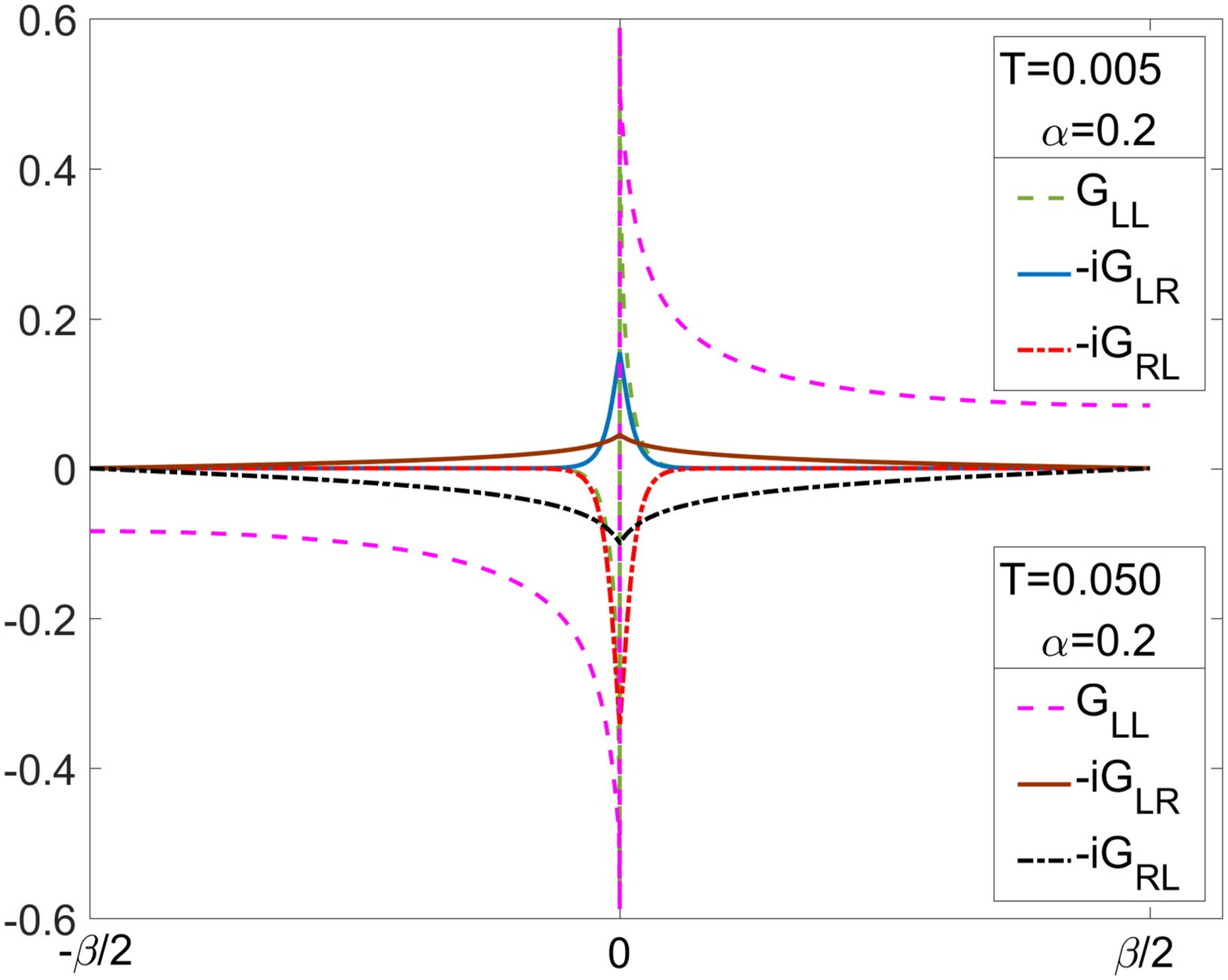}}
\centerline{(b)}
\end{minipage}
\vfill
\begin{minipage}{0.48\linewidth}
\centerline{\includegraphics[width=4.6cm]{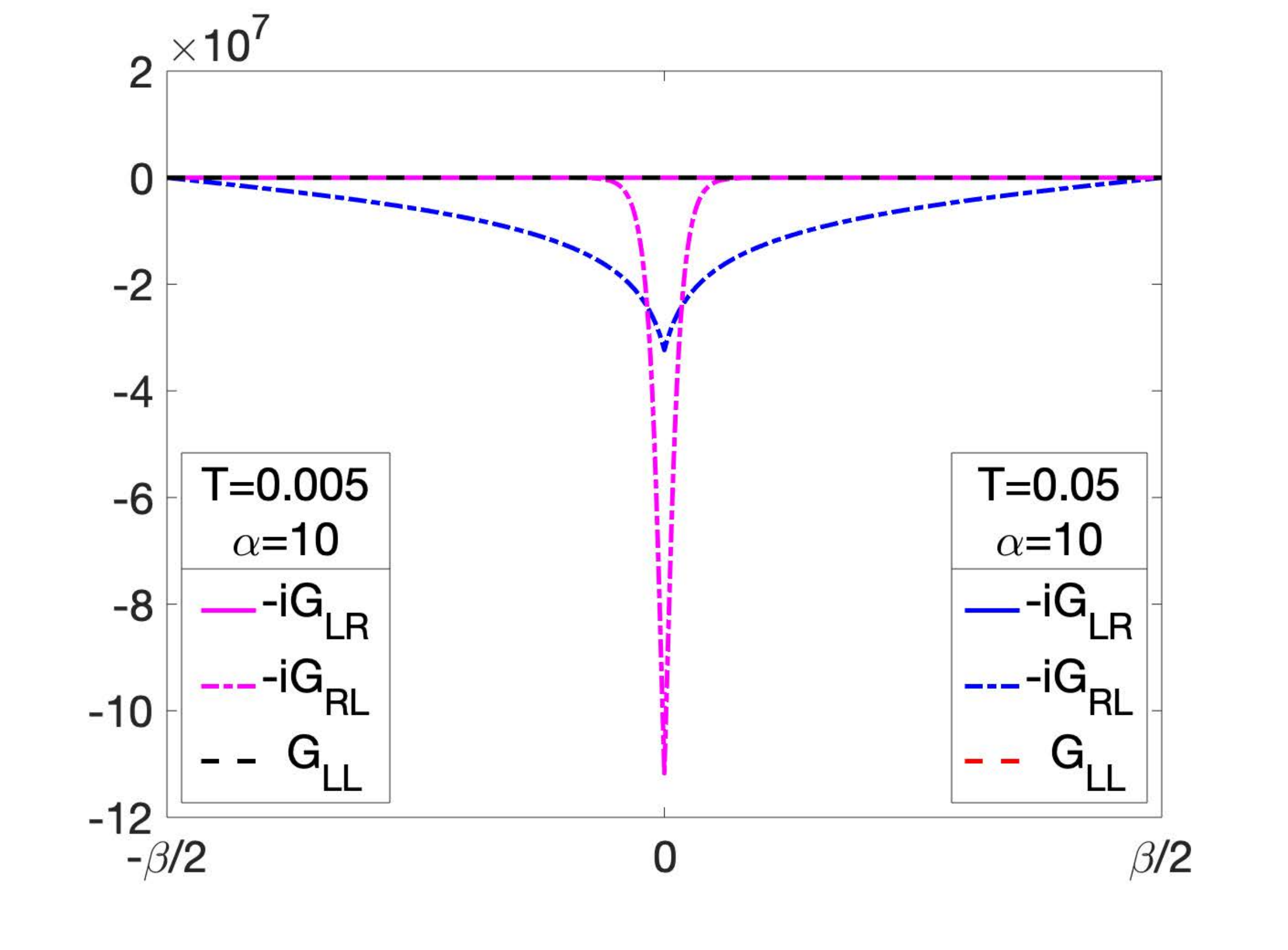}}
\centerline{(c)}
\end{minipage}
\hfill
\begin{minipage}{0.48\linewidth}
\centerline{\includegraphics[width=4.6cm]{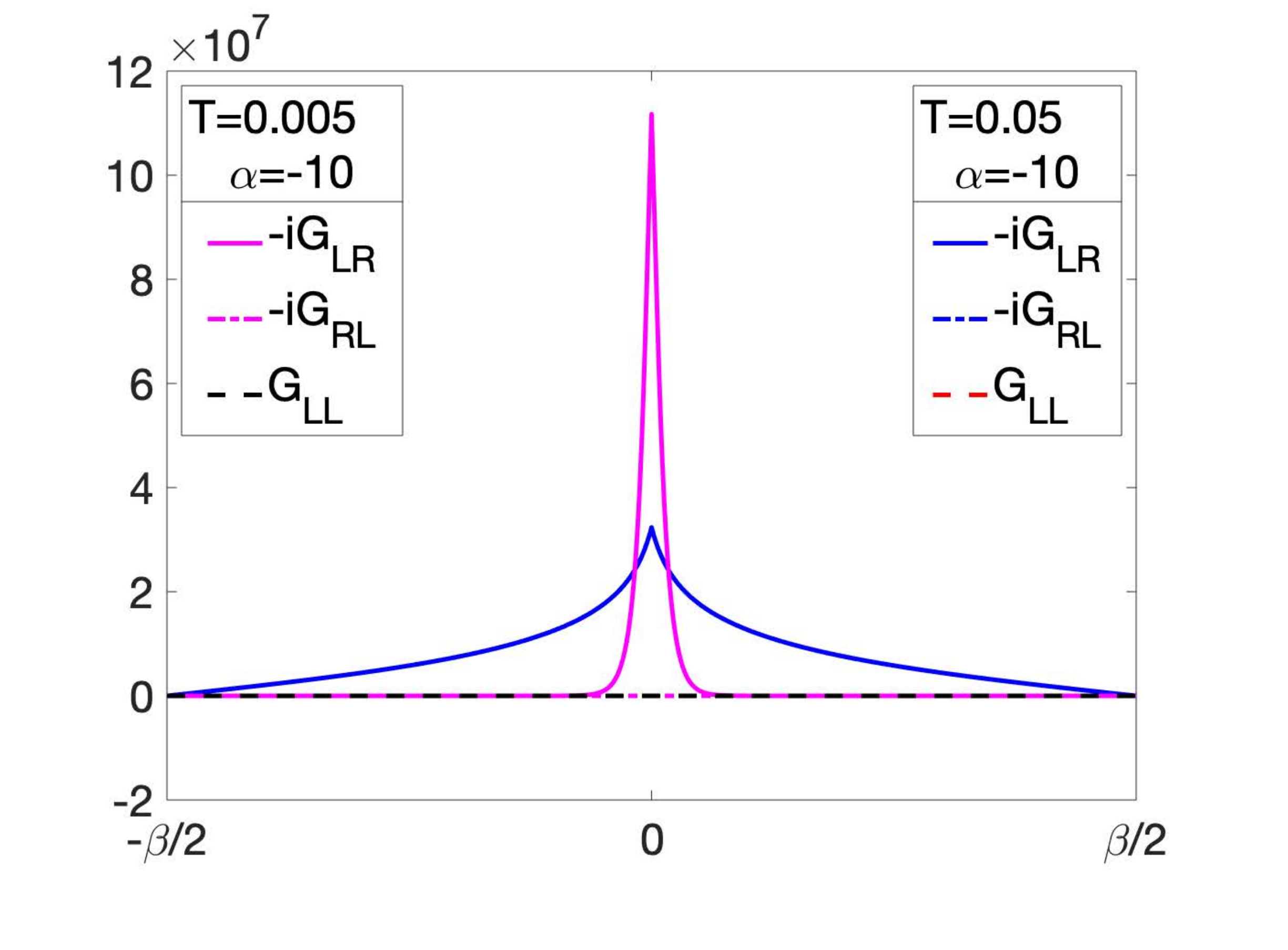}}
\centerline{(d)}
\end{minipage}
\caption{\label{fig:Figure 3} (a) Green's function with $\alpha=0,\pm 0.3$ at temperature $T=0.01$. (b) Green's function with $\alpha=0.2$ at low temperature $T=0.005$ or high temperature $T=0.05$. (c) Green's function with $\alpha=10$ at low temperature $T=0.005$ or high temperature $T=0.05$. (d) Green's function with $\alpha=-10$ at low temperature $T=0.005$ or high temperature $T=0.05$.}
\end{figure}
The numerical results in Fig. \ref{fig:Figure 3}(a)  show that $G_{RR}(i\omega_n,\alpha)=G_{LL}(i\omega_n,\alpha)$, $G_{LR}(i\omega_n,\alpha)=-G_{RL}(i\omega_n,-\alpha)$. When $\alpha=0$, the model recovers the pseudo-complex SYK model at zero chemical potential \cite{Sahoo:2020unu}. Green's function decays exponentially
\begin{equation}
G_{ab}(\tau)\sim e^{-E_{gap}\tau}
\end{equation}
(see appendix C for details) within a certain $\alpha$ region at low temperature, and the correlators decay as a power law like SYK behavior at high temperatures, as shown in Fig. \ref{fig:Figure 3}(b). The Green's functions can be considered as the order parameter. When $\alpha$ is large enough, the off-diagonal Green's function decays exponentially no matter at low temperature $T=0.005$ or high temperature $T=0.05$, and $E_{gap}$ decreases as temperature increases from $T=0.005$ to $T=0.05$(see Fig. \ref{fig:Figure 3}(c) and \ref{fig:Figure 3}(d)). According to the approximate behavior of the saddle-point equations Eq. (\ref{eq:SDeq}), as $\alpha\rightarrow+\infty$, the off-diagonal Green's function $G_{RL}\thicksim-\frac{1}{\Sigma_{LR}}$ dominates while as $\alpha\rightarrow-\infty$, the term $G_{LR}\thicksim-\frac{1}{\Sigma_{RL}}$ dominates. The approximate solutions are indicative of decoupled SYK behavior in the IR limit ($G\thicksim-\frac{1}{\Sigma}$). The results with $\alpha=10$ support this statement numerically in Fig. \ref{fig:Figure 3}(c) and \ref{fig:Figure 3}(d).

We evaluate the free energy of this non-Hermitian coupled model in this section. Substituting the saddle-point solutions into the action in Eq. (\ref{eq:Seff}) as the method in \cite{Cao:2021upq}, we obtain the free energy
\begin{align}
  \frac{F}{N}=&-T\frac{\log Z}{N}=T\frac{S_{eff}}{N}\nonumber\\
   =&-T\bigg[2\log 2+\sum_{\omega_n}\log\frac{D(i\omega_n,\alpha)}{(i\omega_n)^2}+\sum_{\omega_n}\bigg(\frac{3}{4}\Sigma_{LL}(i\omega_n,\alpha)\nonumber\\
   &G_{LL}(i\omega_n,\alpha)+\frac{3}{4}\Sigma_{RR}(i\omega_n,\alpha)G_{RR}(i\omega_n,\alpha)+\frac{3}{4}\Sigma_{LR}(i\omega_n,\alpha)\nonumber\\
   &G_{RL}(i\omega_n,\alpha)+\frac{3}{4}\Sigma_{RL}(i\omega_n,\alpha)G_{LR}(i\omega_n,\alpha)\bigg)\bigg]\,.\label{eq:FE}
\end{align}

\begin{figure}[!t]
\begin{minipage}{0.48\linewidth}
\centerline{\includegraphics[width=4.5cm]{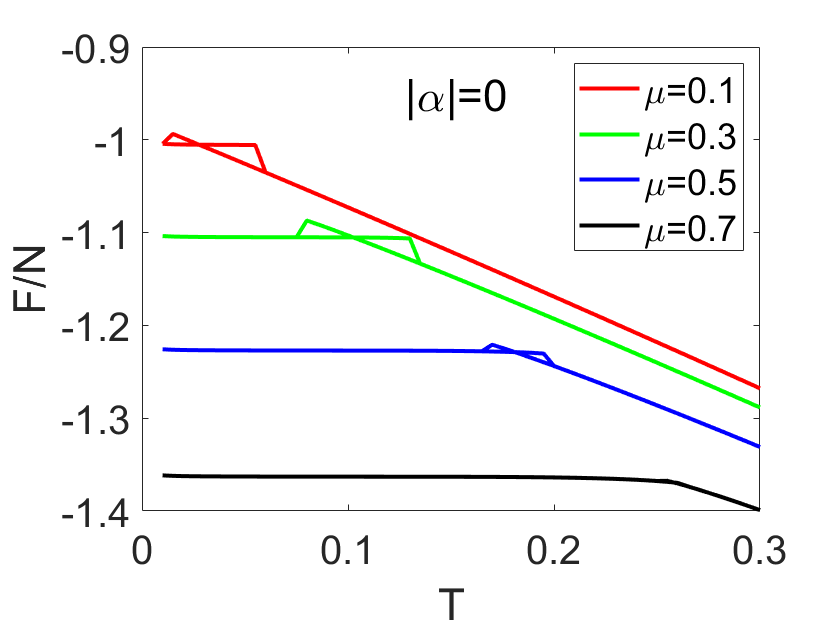}}
\centerline{(a)}
\end{minipage}
\hfill
\begin{minipage}{0.48\linewidth}
\centerline{\includegraphics[width=4.5cm]{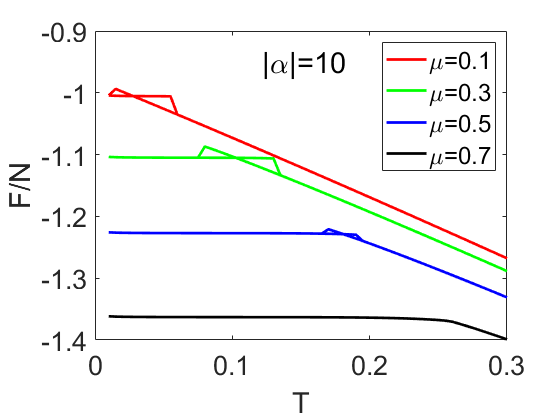}}
\centerline{(b)}
\end{minipage}
\caption{\label{fig:Figure 5} The free energy as a function of $T$ with different coupling $\mu$ for the values of $|\alpha|=0,10$ and $J=1/6$. Starting from the high temperature, we decrease the temperature to low value, and then increase it again back to the high temperature value.}
\end{figure}

The free energy as a function of temperature are plotted in Fig. \ref{fig:Figure 5}.
The free energy obtained in Fig. \ref{fig:Figure 5} is analogous to the free energy of the pseudo-complex SYK model with the Hermitian coupled term in Ref.\cite{Sahoo:2020unu,Ferrari:2019ogc,Garcia-Garcia:2020vyr}.
The first-order phase transition from the low temperature traversable wormhole phase to the high temperature two black hole phase ends at a second-order critical point $(T_c=0.25, \mu_c=0.7)$, which is not influenced by the non-Hermitian parameter.

\section{VI.Conclusion and discussion}
We have constructed a novel non-Hermitian two coupled SYK model yielding a real energy spectrum. This is a pseudo-Hermitian Hamiltonian, where the non-Hermiticity is reflected in the coupling between two copies of the pseudo-complex SYK model. The ground states of the total Hamiltonian receive contribution from the non-Hermitian parameter. In the strong non-Hermitian limit, the wave function $|\psi_0^r\rangle$ approaches the ``ground states'' on the left and ``excited states'' on the right or vice versa. The effective action and the free energy are $\alpha$-independence, although the left and right side states are actually $\alpha$-dependent.  To understand this, let us analyze the non-Hermitian parameter dependence of the free energy. As mentioned previously, only $G_{LR}\thicksim-\frac{1}{\Sigma_{RL}}$ exists in Eq.(\ref{eq:FE}) when $\alpha\rightarrow-\infty$, and only $G_{RL}\thicksim-\frac{1}{\Sigma_{LR}}$ exists in Eq.(\ref{eq:FE}) when $\alpha\rightarrow+\infty$. Since Green's function and the self-energy satisfy $\Sigma_{LR}(i\omega_n,\alpha)=-\Sigma_{RL}(i\omega_n,-\alpha)$, the free energy does not change in the limit $\alpha=\pm\infty$ obviously. Low energy analysis further reveals an $\alpha$-independent energy gap.
However, a key observation is that the off-diagonal Green's functions $G_{LR}$ and $G_{RL}$ are no longer symmetric and strongly $\alpha$-dependent. Note that the transmission amplitude of particles across the traversable wormhole is proportional to the retarded Green's function. Thus, this may elucidate the observable aspects of the dynamics.

\section*{Acknowledgements}
We would like to thank Matteo Baggioli, A.~M.~Garc\'\i{}a-Garc\'\i{}a and Su-Peng Kou for valuable comments and discussions. This work is supported by NSFC China (Grants No.12275166, No. 11875184 and No. 11805117), and Mid-career Researcher Program through the National Research Foundation of Korea Grant no. NRF-
2021R1A2B5B02002603.

  \appendix
\section{A. non-Hermitican similarity transformation}
\setcounter{equation}{0}
\renewcommand\theequation{A.\arabic{equation}}
In this section, we provide more details about a similarity transformation on the left hand-side and the right hand-side Majorana fermion. One can generalize the phase $\phi$ from a real number to an imaginary number $\phi=-i \alpha$ and the imaginary phase transformation becomes a non-Hermitian particle-hole similarity transformation $C\rightarrow e^{\pm\alpha}C$ and $C^{\dagger}\rightarrow e^{\pm\alpha}C^{\dagger}$. It can be written as
\begin{eqnarray}
\psi^L_{2i-1}=e^{\alpha}C^L_i+e^{-\alpha}C^{L\dag}_i\,&,&\,\psi^L_{2i}=i(e^{\alpha}C^L_i-e^{-\alpha}C^{L\dag}_i)\nonumber\\
\psi^R_{2i-1}=e^{-\alpha}C^R_i+e^{\alpha}C^{R\dag}_i\,&,&\psi^R_{2i}=i(e^{-\alpha}C^R_i-e^{\alpha}C^{R\dag}_i)\,.\nonumber\\
\end{eqnarray}
After the self-similarity transformation, the Majorana fermions become non-Hermitian $(\psi^L)^{\dag} \neq (\psi^L)$, $(\psi^R)^\dag\neq\psi^R$ as long as $\alpha \neq 0$ in which the fermions $C^L,C^R,C^{L\dag},C^{R\dag}$ satisfy the anti-commutation relations as follow
\begin{equation}
\{C^{A\dag}_i,C^{B}_j\}=\delta_{ij}\delta_{AB}\,,\,\{C^{A\dag}_i,C^{B\dag}_j\}=\{C^{A}_i,C^{B}_j\}=0\,.\nonumber
\end{equation}
We consider the $\psi_{2i-1}$ case for simplicity of our calculation. In principle, the parameter $\alpha$ could be introduced by performing the non-Hermitican particle-hole similarity transformation on the MQ model
\begin{align}
 H=&-\sum^N_{ijkl}J_{ijkl}(\psi^L_{2i-1}\psi^L_{2j-1}\psi^L_{2k-1}\psi^L_{2l-1}\nonumber\\
   &+\psi^R_{2i-1}\psi^R_{2j-1}\psi^R_{2k-1}\psi^R_{2l-1})+i\mu\sum_i\psi^L_{2i-1}\psi^R_{2i-1}\,.
\end{align}
Note that all the non-physical terms should be dropped out after the similarity transformation. So the Hamiltonian of the $\alpha=0$ case is not exactly that of the MQ model.

\section{B. Noether charge}
\setcounter{equation}{0}
\renewcommand\theequation{B.\arabic{equation}}
In this section, we discuss the Noether charges associated with the $SL(2)$ symmetry in the low energy effective action.
The action is given by the combination of (\ref{eq:SA}) and (\ref{eq:Sint1}) in the main text.
The general form of the Noether charge is described by
\begin{equation}
	Q = \sum_{k \geq 1}^{N} \left( \frac{d}{d \tau} \right)^{k-1} \left[ Y \sum_{m\geq k}^{N} \binom{m}{k}
	 \left(-\frac{d}{d\tau} \right)^{m-k} \frac{\partial L}{\partial h^{(m)}}\right],
\end{equation}
where the infinitesimal transformation is given by $\tau \to \tau + a Y$ with a small parameter $a$.
Since the effective Lagrangian contains up to third-derivative terms, the Noether charge can be explicitly written as
\begin{align}
	Q = Y &\left[ \frac{\partial L}{\partial h'} - \left(\frac{\partial L}{\partial h''} \right)' + \left( \frac{\partial L}{\partial h'''}\right)'' \right] \nonumber\\
	&+ \frac{d Y}{d \tau} \left[ \frac{\partial L}{\partial h''} - 3\left(\frac{\partial L}{\partial h'''} \right)' \right] + \frac{d^{2} Y}{d\tau^2}\left( \frac{\partial L}{\partial h'''} \right).
\end{align}
Considering the $SL(2)$ transformation generated by (\ref{eq:generator}) in the main text, the corresponding charges are given by
\begin{align}
	\frac{Q_{0}}{N} &= Q_{0}[h_{L},\phi_L] + Q_{0}[h_{R},\phi_R]  + \left( \frac{1}{h_{L}'} + \frac{1}{h_{R}'} \right)I_{c}, \\
	\frac{Q_{+}}{N}& = Q_{+}[h_{L},\phi_L]  - Q_{+}[h_{R},\phi_R]   + \left( \frac{e^{ih_{L}}}{h_{L}'} - \frac{e^{ih_{R}}}{h_{R}'} \right)I_{c},\\
	\frac{Q_{-}}{N} &= Q_{-}[h_{L},\phi_L]  - Q_{-}[h_{R},\phi_R]   + \left( \frac{e^{-ih_{L}}}{h_{L}'} - \frac{e^{-ih_{R}}}{h_{R}'} \right)I_{c},
\end{align}
where
\begin{align}
	Q_{0}[h,\phi] &= i \varepsilon K \left( \phi' + i \varepsilon h' \right) -\alpha_{S}\left( h' +\frac{h'''}{h'^{2}} - \frac{h''^{2}}{h'^{3}} \right), \\
	Q_{+}[h,\phi] &= e^{i h}\left[ i \varepsilon K \left( \phi' + i \varepsilon h' \right) -\alpha_{S}\left( -\frac{h''}{h'} +\frac{h'''}{h'^{2}} - \frac{h''^{2}}{h'^{3}} \right) \right], \\
	Q_{-}[h,\phi] &= e^{-i h}\left[ i \varepsilon K \left( \phi' + i \varepsilon h' \right) -\alpha_{S}\left( \frac{h''}{h'} +\frac{h'''}{h'^{2}} - \frac{h''^{2}}{h'^{3}} \right) \right], \\
	I_{c} &=   {\mu \Delta}\bigg[\frac{bh^\prime_Lh^\prime_R}{\cosh^2\frac{h_L-h_R}{2}}\bigg]^\Delta \cosh(\varepsilon h_L-\varepsilon h_R)\nonumber \\
& \times \cosh(2\alpha - i \phi_{L}+i\phi_{R}).
\end{align}
Therefore, the zero charge conditions $Q_{+}=Q_{-}=0$ are satisfied if we set $h_{L}=h_{R}$.
In addition, we study the constraint on the relation between $\phi_{L}$ and $\phi_{R}$ by considering the invariance of the effetive action under the transformation of $h_{a}$ and $\phi_{a}$\,\cite{Garcia-Garcia:2020vyr}.
Then, we require
\begin{align}
	0&=\delta \sum_{A} \left(\phi'_{A} + i\varepsilon h'_{A} \right)^{2} \nonumber\\
	&= 2\sum_{A} \left(\delta \phi'_{A} + i\varepsilon\delta h'_{A} \right) \left(\phi'_{A} + i\varepsilon h'_{A} \right),
\end{align}
and
\begin{align}
	0&=\delta\cosh(2\alpha -i\phi_{L}+i\phi_{R}) \nonumber \\
	&= \sinh(2\alpha -i\phi_{L} + i\phi_{R}) \delta(\phi_{L}-\phi_{R}),
\end{align}
where we have assumed $h_{L}=h_{R}$.
These conditions are satisfied if $\phi_{L}=\phi_{R}-2i\alpha$ and $\delta\phi_{L} = -\delta\phi_{R} = \epsilon(\tau)$, where $\phi_{L}$ and $\phi_{R}$ are constants and $\epsilon(\tau)$ is an arbitrary infinitesimal function.
To be self-consistency, the Noether charge associated with the variation of $\delta\phi_{L} = -\delta\phi_{R} = \epsilon(\tau)$ is given by
\begin{equation}
	Q_{\phi}/N = K \left[ (\phi'_{L}-\phi'_{R}) + i \varepsilon(h'_{L} - h'_{R})\right],
\end{equation}
which vanishes if $h_{L}=h_{R}$ and $\phi_{L}=\phi_{R}-2i\alpha$.

\section{C. energy gap}
\begin{figure}[!t]
\centerline{\includegraphics[width=8.0cm]{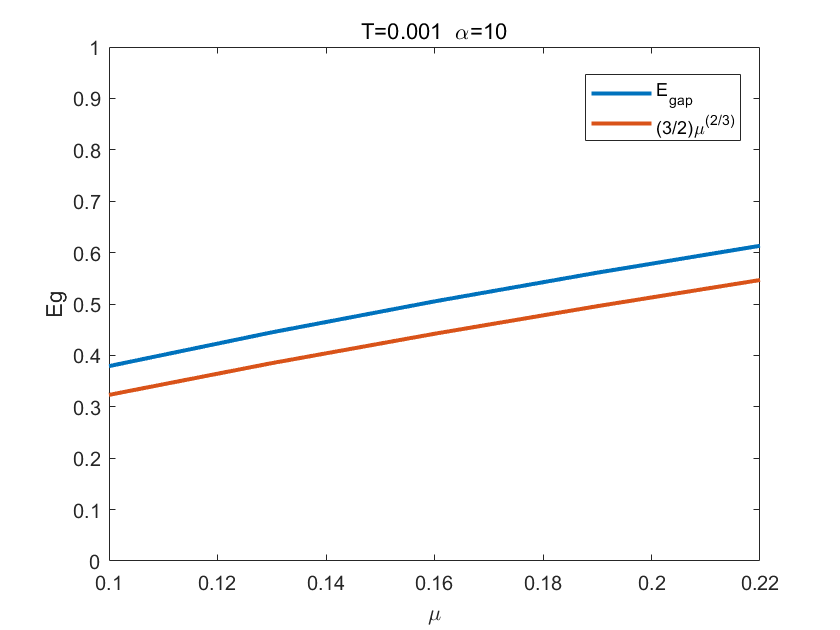}}
\caption{\label{fig:Figure 6} The energy gap extracted from the exponential decay at low temperature $T=0.001$ with $\alpha=10$. The red line represents the power law behavior $E_{gap}\sim\mu^{2/3}$. }
\end{figure}

\setcounter{equation}{0}
\renewcommand\theequation{C.\arabic{equation}}
As mentioned in the main text, Green function decays exponentially $G_{ab}(\tau)\sim e^{-E_{gap}\tau}$. In Fig. \ref{fig:Figure 6}, we show that the energy gap extracted from the exponential decay of $G_{AB}(\tau)$ in which the gap scaling $E_{gap}\sim\mu^{2/3}$, same as that of the Hermitian two coupled SYK model in \cite{Maldacena:2018lmt}.
The gap scaling in Fig. \ref{fig:Figure 6} indicate that our non-Hermitian model return to the Hermitian model consistently if setting $\alpha=0$.

\end{document}